\documentclass{PoS}
\usepackage[utf8]{inputenc}
\usepackage{amssymb}
\usepackage{graphicx}

\title{Some aspects of global Lambda polarization in heavy-ion collisions}

\ShortTitle{Some aspects of global Lambda polarization in heavy-ion collisions}

\author{Xiao-Liang Xia$^a$, Hui Li$^b$, \speaker{Qun Wang}$^c$\\
        Department of Modern Physics, University of Science and Technology of China, Hefei, Anhui 230026, China\\
        \llap{$^a$}E-mail: \email{xiaxl@mail.ustc.edu.cn}\\
        \llap{$^b$}E-mail: \email{lihui12@mail.ustc.edu.cn}\\
        \llap{$^c$}E-mail: \email{qunwang@ustc.edu.cn}}

\abstract{Large orbital angular momentum can be generated in non-central heavy-ion
collisions, and part of it is expected to be converted into final
particle's polarization due to the spin-orbit coupling. Within the
framework of A Multi-Phase Transport (AMPT) model, we studied the
vorticity-induced polarization of $\Lambda$ hyperons at the midrapidity
region $|\eta|<1$ in Au-Au collisions at energies $\sqrt{s_{NN}}=7.7\sim200$
GeV. Our results show that the global polarization decreases with
the collisional energies and is consistent with the recent STAR measurements.
This behavior can be understood by less asymmetry of participant matter
in the midrapidity region due to faster expansion of fireball at higher
energies. As another evidence, we discuss how much the angular momentum
is deposited in different rapidity region. The result supports our
asymmetry argument.}

\FullConference{Critical Point and Onset of Deconfinement - CPOD2017\\
		7-11 August, 2017\\
		The Wang Center, Stony Brook University, Stony Brook, NY}

\begin{document}

\section{Introduction}

In non-central heavy ion collisions, initial orbital angular momentum
as large as $10^{5}\hbar$ can be generated. Part of the angular momentum
is expected to be converted into the polarization of final particles
through spin-orbital coupling \cite{Liang:2004ph,Voloshin:2004ha,Betz:2007kg}.
Using $\Lambda$ hyperon's weak decay, its polarization can be measured
via the angular distribution of the daughter protons \cite{Abelev:2007zk}.
Recently this topic has attracted much attention as STAR collaboration
reported the observation for the global polarization of $\Lambda$($\bar{\Lambda}$)
hyperons in 20\%-50\% centrality Au-Au collisions in the beam energy
scan (BES) program \cite{STAR:2017ckg}, see recent reviews from experimental
\cite{Voloshin:2017kqp} and theoretical perspective \cite{Becattini:2017vsh,Wang:2017jpl}.
The STAR's result shows that the global polarization is significantly
nonzero at low collisional energies and decreases with the collisional
energy in range $\sqrt{s_{NN}}=7.7\sim200$ GeV. This energy dependence
is consistent with the results from hydrodynamic and transport model's
calculations \cite{Karpenko:2016jyx,Xie:2017upb,Li:2017slc,Sun:2017xhx},
but it seems contradictory to the naive expectation that the global
polarization should be proportional to the total orbital angular momentum
and then to the collisional energy. We have studied this problem in
our previous paper \cite{Li:2017slc}, and attributed the polarization's
energy behavior to less asymmetric participant matter in the midrapidity
at higher energies. In this work, we provide further supports of this
conclusion by investigating how much angular momentum is deposited
in different rapidity region.

\section{Model description}

In our study, the string-melting version of A Multi-Phase Transport
(AMPT) model \cite{Lin:2004en} is employed to simulate the evolution
of the collisional system. In the model, all participants in collisions
are converted into partons. By coarse-graining the collective motion
of the partons and adopting the equation of state from lattice result
\cite{Borsanyi:2012cr,Bazavov:2017dus}, we obtain hydrodynamic quantities
such as the fluid velocity $u^{\mu}$, the temperature $T$ and the
thermal vorticity $\varpi_{\mu\nu}$. The thermal vorticity is defined
by
\begin{equation}
\varpi_{\mu\nu}=\frac{1}{2}\left[\partial_{\nu}\left(\frac{u_{\mu}}{T}\right)-\partial_{\mu}\left(\frac{u_{\nu}}{T}\right)\right].\label{eq:thermal_vorticity}
\end{equation}
In local equilibrium, $\Lambda$'s spin four-vector is related to
$\varpi_{\mu\nu}$ by \cite{Becattini:2007nd,Becattini:2013fla,Fang:2016vpj}
\begin{equation}
S^{\mu}=-\frac{1}{8m}\left(1-n_{F}\right)\epsilon^{\mu\nu\rho\sigma}p_{\nu}\varpi_{\rho\sigma},\label{eq:spin}
\end{equation}
where $p^{\mu}$ and $m$ are the four-momentum and mass of the $\Lambda$
hyperon respectively, and $n_{F}=1/[1+\exp(\beta\cdot p\mp\mu/T)]$
is the Fermi-Dirac distribution function for particle (upper sign)
and anti-particle (lower sign). At the end of the partonic stage in
the AMPT model, all partons are converted into hadrons. Then we collect
the $\Lambda$($\bar{\Lambda}$) hyperons in the midrapidity region
$|\eta|<1$ and calculate their spin vectors by Eq.~(\ref{eq:spin}).
In the center of mass frame (c.m.f.) of two nuclei, the spin four-vector
$S^{\mu}=(S^{0},\mathbf{S})$ has a non-zero time component $S^{0}$.
To obtain the spin three-vector, we should boost $S^{\mu}$ into the
$\Lambda$'s rest frame, where the spin four-vector $S^{\mu*}$ does
not have the time component, i.e.~$S^{\mu*}=(0,\mathbf{S}^{*})$
with $\mathbf{S}^{*}$ being the spin three-vector and related to
$\mathbf{S}$ by
\begin{equation}
\mathbf{S}^{*}=\mathbf{S}-\frac{\mathbf{p}\cdot\mathbf{S}}{E_{p}\left(m+E_{p}\right)}\mathbf{p},
\end{equation}
where $E_{p}$ and $\mathbf{p}$ are the $\Lambda$'s energy and momentum
in the c.m.f.~of two nuclei. The global $\Lambda$ polarization is
obtained by $P=2\left\langle \mathbf{S}^{*}\right\rangle \cdot\hat{\mathbf{J}}$,
where $\hat{\mathbf{J}}$ is the unit vector along the direction of
the collisional system's orbital angular momentum (perpendicular to
the reaction plane), the average is taken over all $\Lambda$ hyperons,
and we have introduced a normalization factor 2 (since the $\Lambda$'s
spin is 1/2). In the AMPT's coordinate system, the $x$-axis is along
the direction of impact parameter $\mathbf{b}$ and $z$ is along
the beam direction. Two nuclei centered at $(x=\pm b/2,\ y=0)$ move
along the $\pm z$ direction, respectively. Therefore $\hat{\mathbf{J}}$
is along $-y$ direction and the global polarization is simply $P=-2\left\langle S_{y}^{*}\right\rangle $.

\section{Numerical results from AMPT}

\subsection{The global $\Lambda$ polarization}

We run the simulations at collisional energies $\sqrt{s_{NN}}=7.7$,
11.5, 14.5, 19.6, 27, 39, 62.4 and 200 GeV. For each energy, we chose
two specific impact parameters $b=7$ and 9 fm corresponding to 20\%-50\%
centrality class in the STAR experiment. The final results are obtained
by taking average over these two impact parameters.

\begin{figure}
\centering\includegraphics[width=0.5\linewidth]{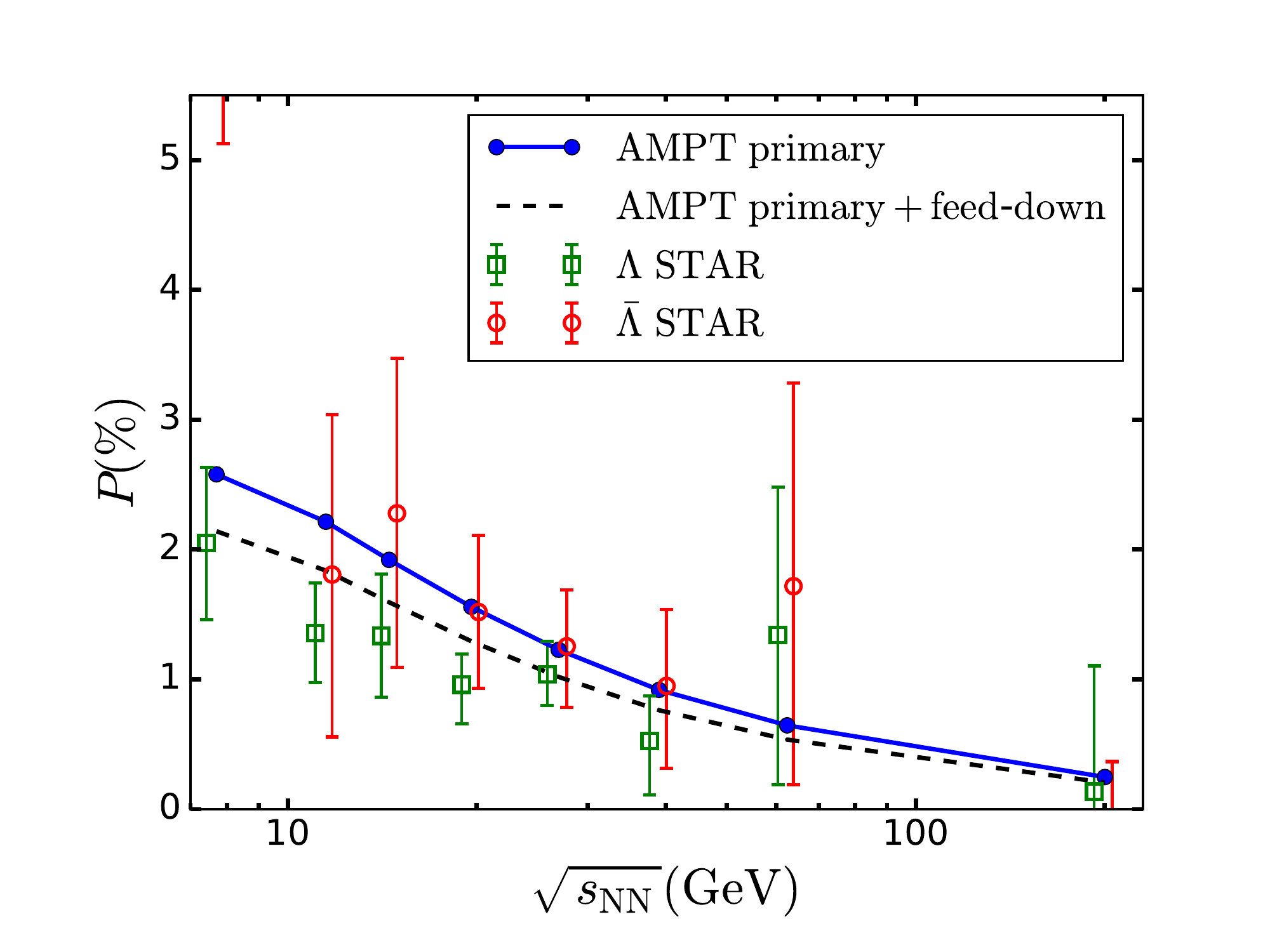}

\caption{\label{fig:polarization}The energy dependence of the global $\Lambda$
polarization in the midrapidity region $|\eta|<1$ from $7.7$ to
$200$ GeV. The curves represent the global polarization of primary
$\Lambda$s (blue solid) and primary plus feed-down $\Lambda$s (black
dashed). The unfilled squares and circles represent the STAR data
for the global $\Lambda$ and $\bar{\Lambda}$ polarization \cite{STAR:2017ckg}.}
\end{figure}

Fig.~\ref{fig:polarization} shows the energy dependence of the global
$\Lambda$ polarization in the midrapidity region $|\eta|<1$. The
blue solid line counts only the $\Lambda$ hyperons produced directly
from quark coalescence (known as primary $\Lambda$). In reality,
there are also secondary $\Lambda$s produced from the resonance decay
(feed-down). These feed-down $\Lambda$s are estimated to suppress
the global polarization by about 17\% from the primary-only result
\cite{Becattini:2016gvu,Karpenko:2016jyx,Li:2017slc}. The global
polarization of primary plus feed-down $\Lambda$s (black dashed line)
is in agreement with the STAR's result.

We see that the global $\Lambda$ polarization decreases with the
collisional energy: from about 2.1\% at 7.7 GeV to 0.2\% at 200 GeV.
By investigating the spatial distributions of the thermal vorticity
and the $\Lambda$ hyperon's position, we give the explanation on
this energy behavior in our previous paper \cite{Li:2017slc}. In
non-central collisions, the fireball is tilted on the reaction plane
($x$-$z$ plane) after collisions: there is more matter in $xz>0$
region than in $xz<0$ region. Such an asymmetry plays a crucial role
in generating a net vorticity and the global polarization along the
direction of the collisional system's angular momentum ($-y$ direction),
see Ref.~\cite{Li:2017slc} for details. However at high energy such
as 200 GeV, the fireball extends over a wider rapidity range due to
its fast longitudinal expansion. As a consequence, the tilted shape
of the fireball is only visible at large rapidity, while at midrapidity
the matter is almost symmetrically distributed and thus the global
polarization nearly vanishes.

\subsection{Initial parton distribution and angular momentum}

\begin{figure}
\centering\includegraphics[width=0.33\linewidth]{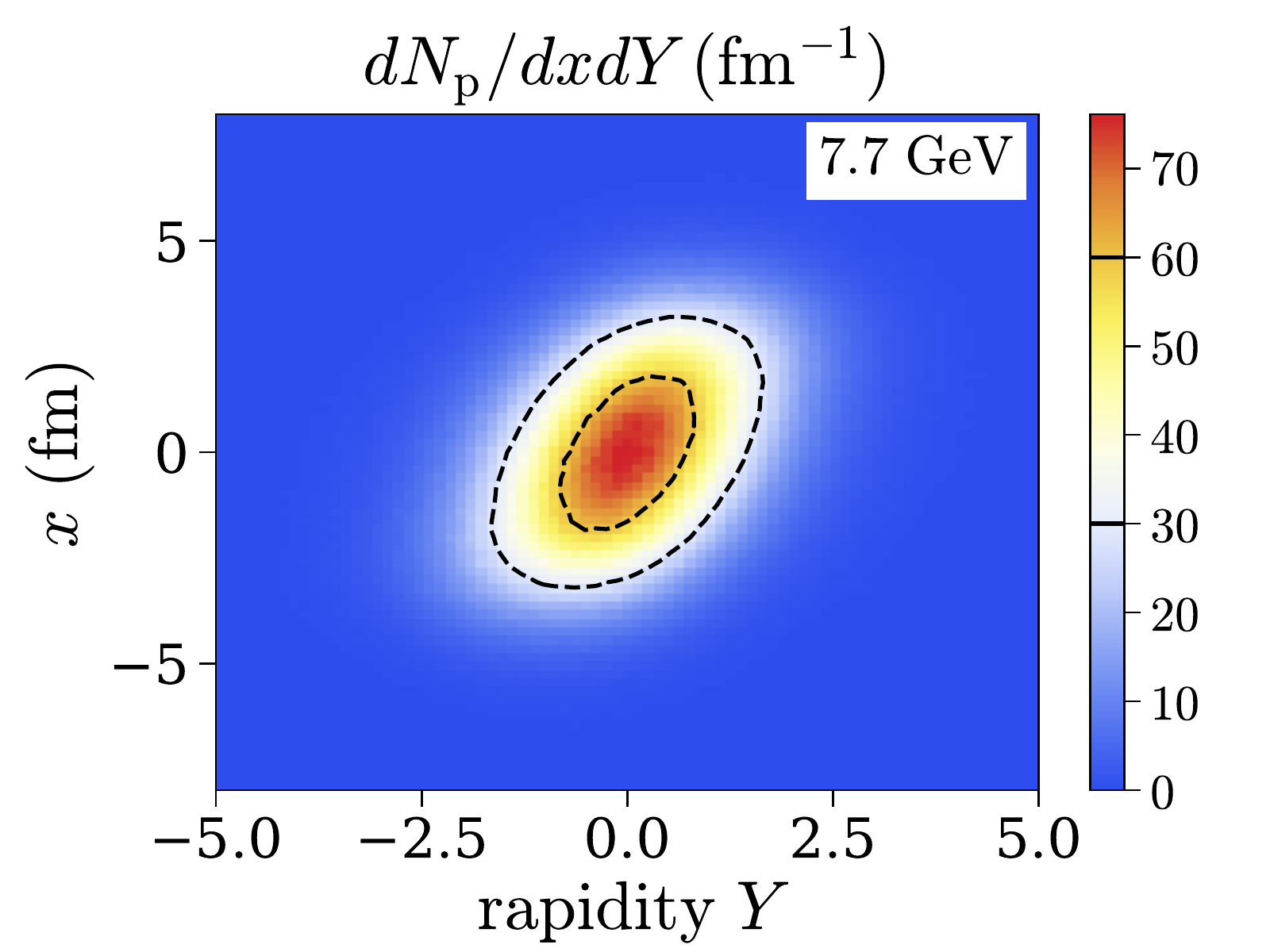}\includegraphics[width=0.33\linewidth]{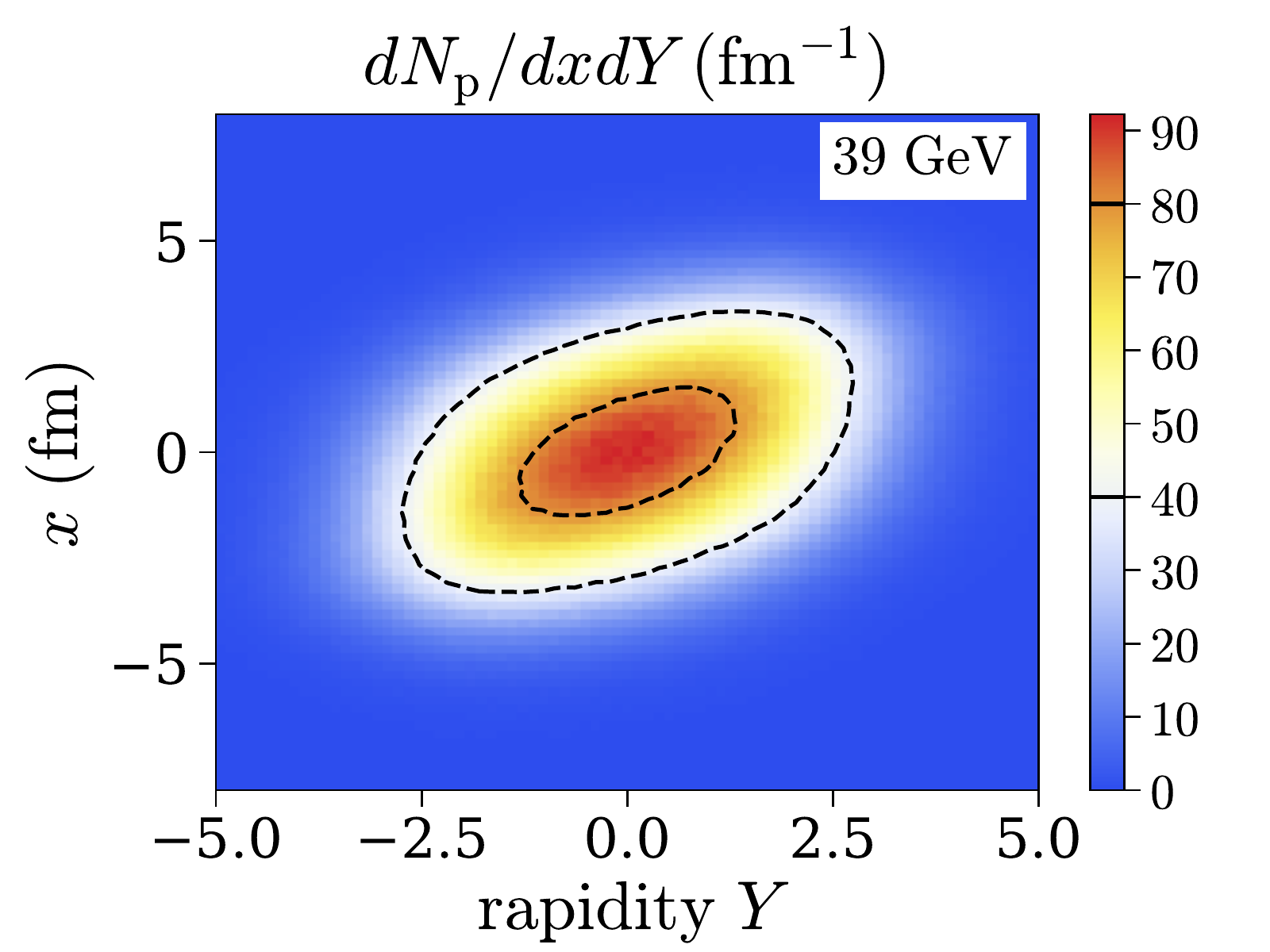}\includegraphics[width=0.33\linewidth]{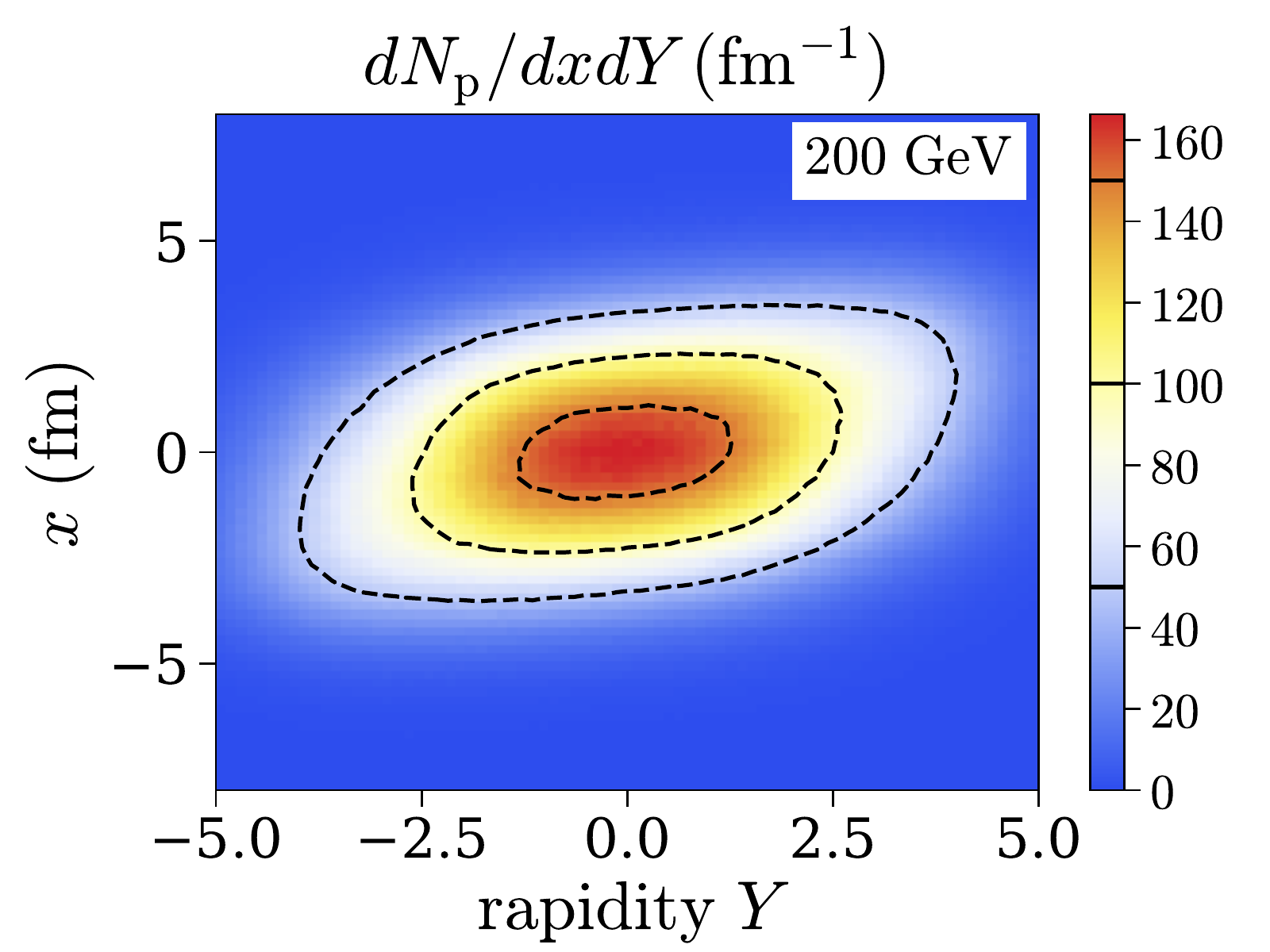}

\caption{\label{fig:parton distribution}The initial parton distribution as
functions of the transverse coordinate $x$ and longitudinal rapidity
$Y$ at collisional energies $\sqrt{s_{NN}}=7.7$ GeV (left), 39 GeV
(middle) and 200 GeV (right) and the impact parameter $b=7$ fm. For
better illustration, we draw equal density contours (black dashed
lines) in each panel.}
\end{figure}

We now give an illustration of the fireball's tilted shape. For simplicity,
we consider the parton distribution at the initial time of the partonic
phase in the AMPT model. At this time, all partons are located at
$z=0$ and are distributed with their transverse positions $(x,y)$
and momenta $(p_{x},p_{y},p_{z})$. Fig.~\ref{fig:parton distribution}
shows the initial parton distribution as functions of the transverse
coordinate $x$ and longitudinal rapidity $Y$, where $Y$ is related
to $p_{z}$ by $Y=\mathrm{arctanh}(p_{z}/E)$. In the figure we take
$\sqrt{s_{NN}}=7.7$, 39, 200 GeV and $b=7$ fm for illustration.
One can see that the parton distributions at all three energies are
tilted on the reaction plane: more partons are in the region $x,Y>0$
or $x,Y<0$ in non-central collisions. Such an asymmetric parton distribution
is the origin of the orbital angular momentum in the system, which
can be calculated by
\begin{equation}
J_{y}=-\sum_{i}x_{i}p_{z}^{i},\label{eq:angular_momentum}
\end{equation}
where the sum is over all partons. One can see from Eq.~(\ref{eq:angular_momentum})
that more partons distributed in the $xp_{z}>0$ region can lead to
a negative net angular momentum after taking the sum.

We also see from Fig.~\ref{fig:parton distribution} that the parton
distribution has a collisional energy and rapidity dependence. At
low energies such as 7.7 GeV, the partons are distributed in smaller
rapidity range with an obvious tilted shape. However at high energies
like 200 GeV, the fireball shows an almost symmetric distribution
at midrapidity, while the tilted shape is only visible at large rapidity
due to the wider extension of the fireball. The collisional energy
and rapidity dependence of parton distributions can lead to the same
dependence of the orbital angular momentum.

\begin{figure}
\centering\includegraphics[width=0.33\linewidth]{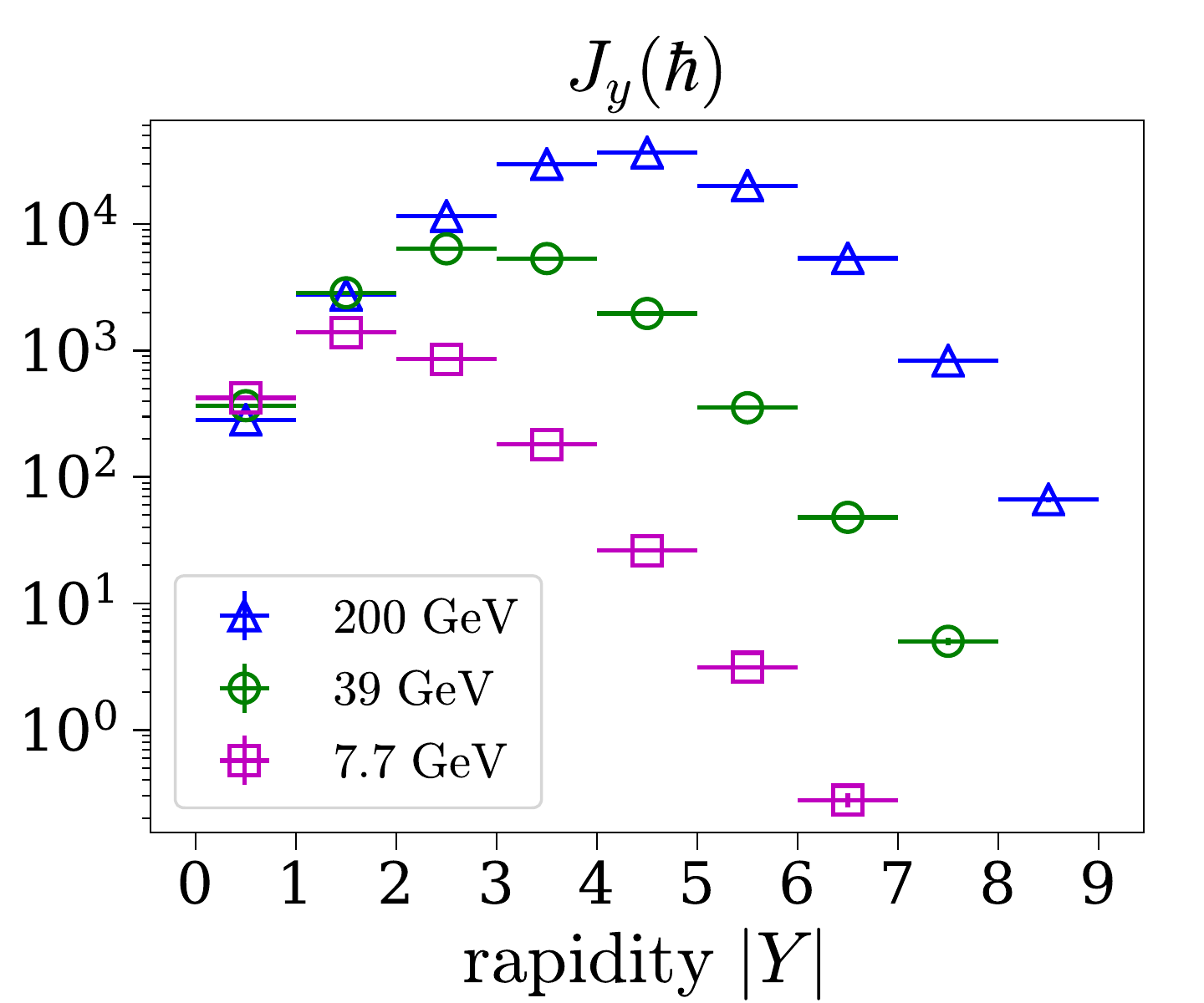}\includegraphics[width=0.33\linewidth]{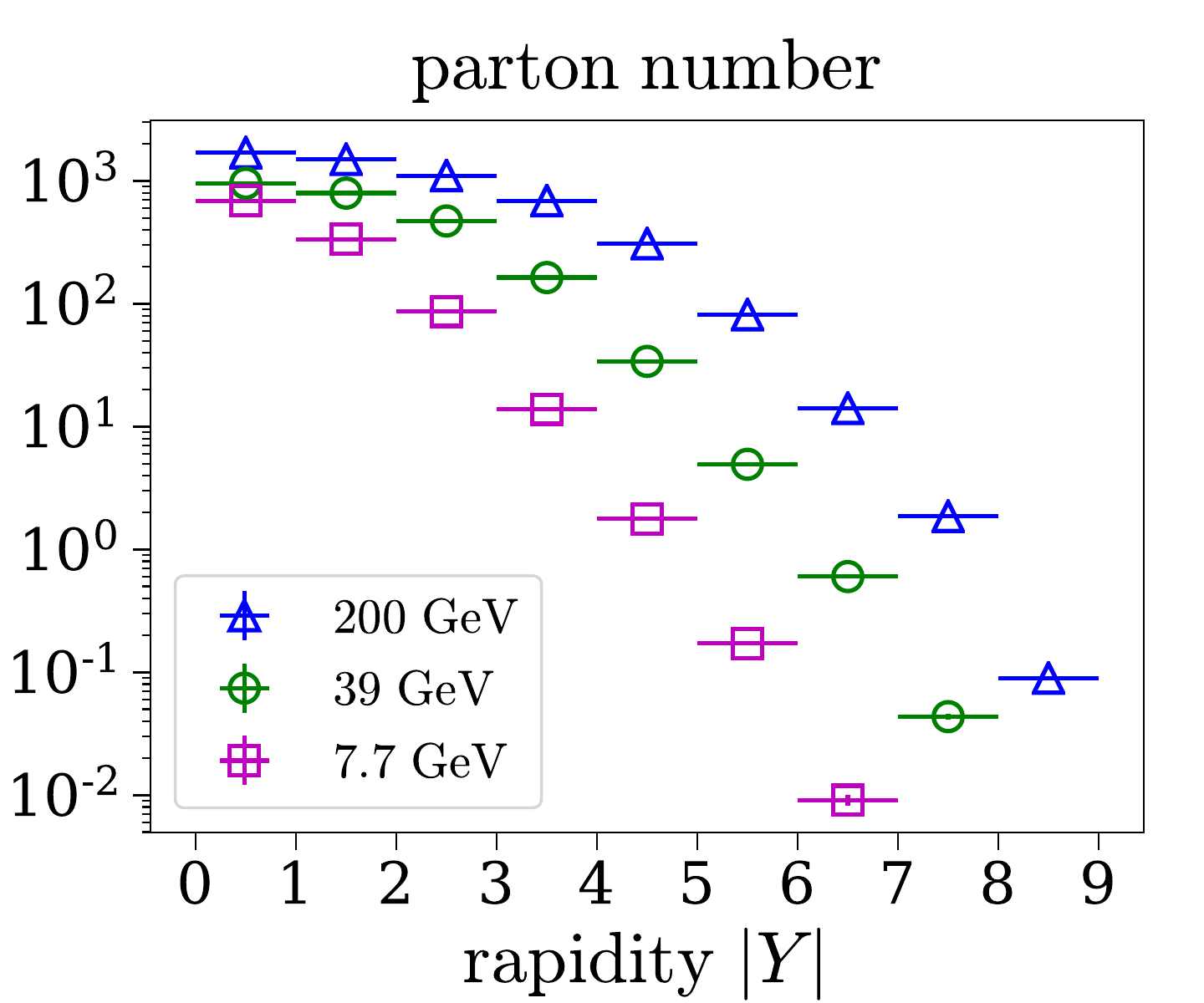}\includegraphics[width=0.33\linewidth]{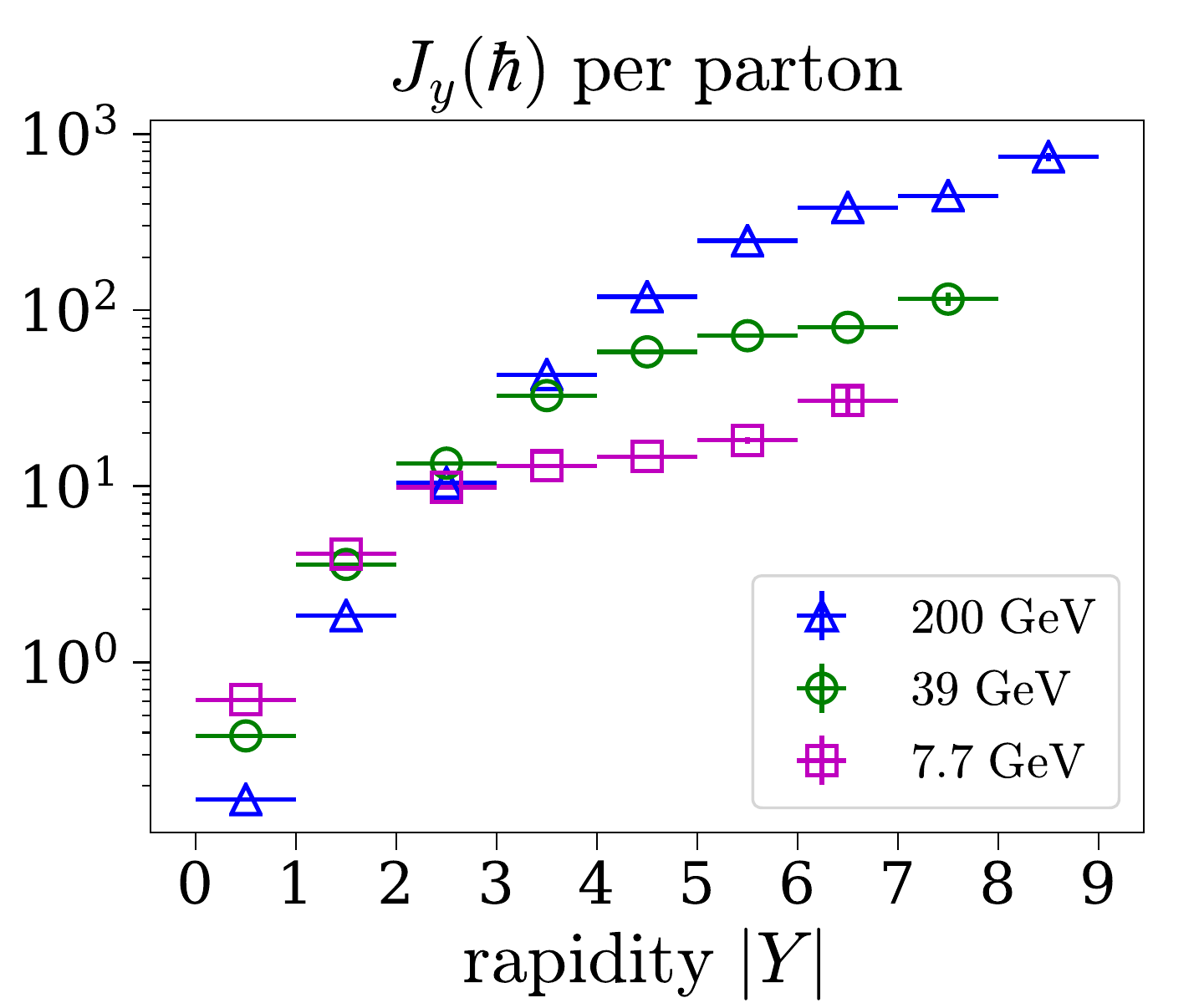}

\caption{\label{fig:angular momentum}The orbital angular momentum $J_{y}$
carried by partons (left), the parton number (middle), and the orbital
angular momentum $J_{y}$ per parton (right) in different rapidity
bins at collisional energies $\sqrt{s_{NN}}=7.7$ GeV (magenta square),
39 GeV (green circle) and 200 GeV (blue triangle) with the impact
parameter $b=7$ fm. Note that in our coordinate system $J_{y}$ is
always negative, but values in the figures are its magnitudes without
the sign.}
\end{figure}

Fig.~\ref{fig:angular momentum} shows how the angular momentum is
deposited in different rapidity bins at three collisional energies,
where the width of the rapidity bin is set to be one unit. For each
collisional energy, $10^{4}$ events are generated for statistics.
In the left panel, the shapes of the angular momentum are similar
for each energy. $J_{y}$ first increases with rapidity due to the
increase of both $|p_{z}|$ and the asymmetry, and then decreases
at larger rapidity. The decrease behavior is because of less partons
at larger rapidity as shown in the middle panel. To remove the effect
of decreasing parton number with rapidity, we divide $J_{y}$ by the
parton number in each bin. The result shown in the right panel turns
out to increase with rapidity monotonically.

\begin{figure}
\centering\includegraphics[width=0.45\linewidth]{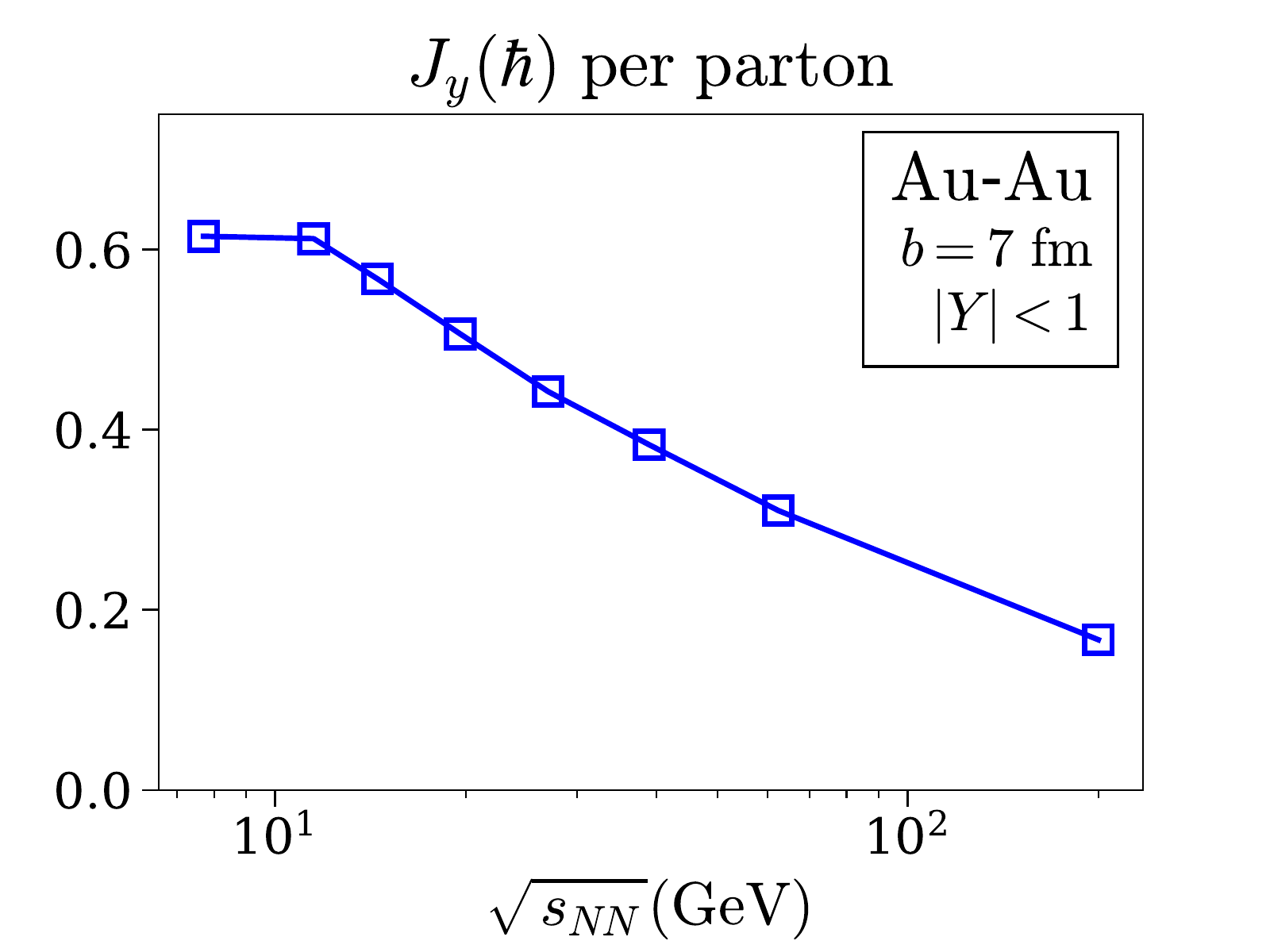}

\caption{\label{fig:angular_mid}The orbital angular momentum $J_{y}$ per
parton in the rapidity bin $|Y|<1$ as a function of the collisional
energy from 7.7 to 200 GeV.}
\end{figure}

Since more partons with larger $|p_{z}|$ are generated in higher
energy collisions, the total angular momentum in the system should
increase with the collisional energy, see Refs.~\cite{Jiang:2016woz,Karpenko:2016jyx}
for example. However, such an energy behavior is not true for the
angular momentum deposited in the midrapidity region. Actually, as
shown in Fig.~\ref{fig:angular momentum}, most of the total angular
momentum is carried away by partons in larger rapidity. In the middle
rapidity bin $|Y|<1$, Fig.~\ref{fig:angular_mid} shows the angular
momentum per parton increases with decreasing collisional energy.
This energy behavior is the same with the energy dependence of the
global polarization \cite{STAR:2017ckg} and the directed flow slope
$dv_{1}/d\eta$ \cite{Voloshin:2017kqp}. It indicates that the hot
and dense matter created in non-central collisions is more tilted
at lower energies in the midrapidity region.

We note that the curve in Fig.~\ref{fig:angular_mid} is almost flat
between 7.7 and 11.5 GeV, although more tilted parton distribution
on the reaction plane is preferred at 7.7 GeV. This can be understood
by that the parton density $dN_{p}/dY$ sharply decreases with rapidity
at 7.7 GeV, making the mean value of $|p_{z}|$ smaller than that
at higher energies.

\section{Summary}

We have studied the vorticity-induced polarization of the $\Lambda$
hyperons in the midrapidity region for Au-Au collisions at $\sqrt{s_{NN}}=7.7\sim200$
GeV with the AMPT model. Our result is in good agreement with the
recent STAR measurement. The value of global $\Lambda$ polarization
at 7.7 GeV is about one order of magnitude larger than that of 200
GeV. We also investigate how the angular momentum is distributed in
the system. The result shows that although the angular momentum in
the whole system increases with collisional energy, the one deposited
in the midrapidity per parton decreases on the contrary. The latter
energy behavior is the same with the global polarization, and indicates
the participant matter in non-central collisions is more tilted at
lower energies in the midrapidity.

\section*{Acknowledgements}

The authors are supported in part by the Major State Basic Research
Development Program (973 Program) in China under the Grant No.~2015CB856902
and 2014CB845402 and by the National Natural Science Foundation of
China (NSFC) under the Grant No.~11535012.

\bibliographystyle{JHEP}
\bibliography{ref}

\end{document}